\documentstyle[aps,prl,twocolumn, epsf]{revtex}
\def\be{\begin{equation}}
\def\ee{\end{equation}}
\def\bea{\begin{eqnarray}}
\def\eea{\end{eqnarray}}
\def\bma{\begin{mathletters}}
\def\ema{\end{mathletters}}
\def\P{{\cal P}}
\def\C{{\cal C}}
\def\E{{\cal E}}
\def\R{{\cal R}}
\def\w{{\mbox w}}

\newcommand{\bra}[1]{\mbox{$\langle #1 |$}}
\newcommand{\ket}[1]{\mbox{$| #1 \rangle$}}
\newcommand{\braket}[2]{\mbox{$\langle #1  | #2 \rangle$}}
\newcommand{\proj}[1]{\ket{#1}\!\bra{#1}}

\begin{document}
         
\draft

\title{Storage of quantum dynamics in quantum states: \\
a quasi-perfect programmable quantum gate}

\author{G. Vidal and J. I. Cirac}

\address
{Institut f\"ur Theoretische Physik, Universit\"at Innsbruck,A-6020 Innsbruck, Austria}

\date{\today}

\maketitle

\begin{abstract}
We show how quantum dynamics (a unitary transformation) can be captured in the state of a quantum system, in such a way that the system can be used to perform, at a later time, the stored transformation almost perfectly on some other quantum system. Thus programmable quantum gates for quantum information processing are feasible if some small degree of imperfection is allowed.  We discuss the possibility of using this fact for securely computing a secret function on a public quantum computer. Finally, our scheme for storage of operations also allows for a new form of quantum remote control.

\end{abstract}


\bigskip

\section{Introduction}

Quantum information theory explores the potential of quantum mechanics in order to process and transmit information. A two-level system, a qubit, constitutes the unit resource for storing information. Similarly, a unitary operation on one qubit can be regarded as a basic unit of information processing \cite{explanation}.
 In this paper we explore the possibility of storing quantum dynamics, in particular unitary transformations, in the state of a quantum system, in a manner that the transformation can be performed at a later time and on another system almost perfectly. 

\subsection{Quantum programmable gates.}

 The problem we address can be well-posed in the context of quantum circuitry. We will say that the {\em program} state $\P_U$ of some {\em program register} stores the one-qubit transformation $U$, if some ``fixed'' protocol employing the state $\P_U$ is able to perform $U$ on an arbitrary {\em data} state $\rho$ of a single qubit {\em data register}. Here, a ``fixed'' protocol means that the manipulation of the joint state 
\be
\rho\otimes\P_U 
\label{estat}
\ee
does not require knowing the operation $U$ nor the state $\rho$. A device able to transform state (\ref{estat}) into 
 \be
U\rho U^{\dagger}\otimes \R_{\rho,U},
\label{universal}
\ee
where $\R_{\rho,U}$ is just some residual state, is known as a {\em programmable} quantum gate \cite{Nie97}. In a similar fashion as modern (classical) computers take both the program to be executed and the data to be processed as input bit strings, a programmable or universal quantum gate is a device whose action $U$ on an arbitrary data state $\rho$ is determined by the program state $\P_U$. 

 Nielsen and Chuang \cite{Nie97} analyzed the possibility of constructing such a programmable quantum gate. Its total dynamics was described by means of a fixed unitary operator G according to
\be
G[\ket{d}\otimes\ket{\P_U}] = (U\ket{d})\otimes\ket{\R_U},
\label{perfect}
\ee
where only pure data states $\ket{d}$ were considered because this already warranties the mixed state case. Notice that the program state $\ket{\P_U}$ and the residual state $\ket{\R_U}$ --- which was showed to be independent of $\ket{d}$--- can always be taken to be pure, by extending the program register with an ancillary system if needed.
Nielsen and Chuang proved that any two inequivalent operations $U$ and $V$ require orthogonal program states, that is $\braket{\P_U}{\P_V}=0$.
Thus, in order to perfectly store a given operation $U_i$ from some set $\{U_i\}_{i\in I}$, a vector state $\ket{\P_{U_{i}}}$ from an orthonormal basis $\{\ket{\P_{U_{i}}}\}_{i\in I}$ has to be used. The operation $U_{i}$ can then be implemented by, say, measuring the program register to obtain the value $i$, and gauging correspondingly some convenient experimental device.
 Since the set of unitary operations is infinite, their result implied that no universal gate can be constructed using finite resources, that is, with a finite dimensional program register.

\subsection{Main results.}

 The aim of this work is to present programmable quantum gates with a finite program register, and thus physically feasible. A finite register turns out to be sufficient if a degree of imperfection, no matter how small, is allowed in performing the unkwon operation $U$. We will construct a family of {\em probabilistic} programmable quantum gates, that is programmable quantum gates which work with a given prior probability $p \geq 1-\epsilon$ of a successful implementation of $U$. Such a one-qubit gate with $\epsilon = 3/4$ was already described in \cite{Nie97}. Here we will achieve any arbitrarily small $\epsilon > 0$. We will also consider {\em approximate} programmable quantum gates, which perform an operation $\E_U$ very similar to the desired $U$, that is $F(\E_U,U) \geq 1-\epsilon$ for some transformation fidelity $F$. 

The second main result is a lower bound on the dimension of the program register of the programmable gate in terms of its degree of imperfection $\epsilon$. It implies that the orthogonality result of \cite{Nie97} is robust. We will discuss its implications in the context of secure secret computation. 

Finally, operations stored in a quantum state can be teleported. This leads to a new scheme for quantum remote control \cite{Hue00} that only requires unidirectional communication.

\section{Quasi-Perfect programmable quantum gates}

 We start by showing how to store and reimplement, in an imperfect but feasible fashion, an arbitrary one-qubit unitary operation of the form
\be
U_{\alpha} \equiv \exp(i\alpha \sigma_z),
\label{unialpha}
\ee
where $\alpha \in [0,\pi)$. Notice that a general one-qubit operation $U\in SU(2)$ can be obtained by composing three operations of the form of eq. (\ref{unialpha}) with some fixed unitary operations, for instance as $U=U_{\alpha_3}$$\exp(-i\pi\sigma_x/2)$$U_{\alpha_2}\exp(i\pi\sigma_x/2)$$U_{\alpha_1}$. 

\subsection{Single-qubit program state.}

Let us consider the state
\be
\ket{\alpha} \equiv \frac{1}{\sqrt{2}} (e^{i\alpha}\ket{0}+e^{-i\alpha}\ket{1}), 
\ee
which someone, say Alice, can prepare by applying $U_{\alpha}$ on a qubit in the standard state $(\ket{0}+\ket{1})\sqrt{2}$. Suppose she also prepares, along with $\ket{\alpha}$, another qubit in some arbitrary state $\ket{d}=a\ket{0} + b\ket{1}$ and provides Bob, who doesn't know $\alpha$ nor the complex coefficients $a$ and $b$, with the two qubits in state $\ket{d}\otimes\ket{\alpha}$. Alice challenges now Bob to obtain the state $U_{\alpha}\ket{d}$. 

\begin{figure}
 \epsfysize=1.8cm
\begin{center}
 \epsffile{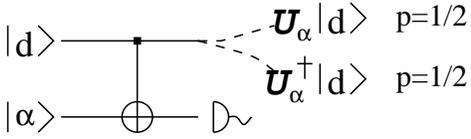}
\end{center}
 \caption{This simple quantum circuit implements a probabilistic gate that takes unknown data and program states $\ket{d}$ and $\ket{\alpha}$ and produces, depending on the result of a measurement on the program register, either $U_{\alpha}\ket{d}$ or $U_{\alpha}^{\dagger}\ket{d}$, with equal prior probability $1/2$ (i.e. $\epsilon = 1/2$).  \label{fig1}}
\end{figure}

 What Bob can do in order to implement  the unknown $U$ with some probability of success is to perform a C-NOT operation taking the data qubit in state $\ket{d}$ as the control and the program qubit in state $\ket{\alpha}$ as the target. This will constitute the basic part of our simplest programmable quantum gate. Recalling that the C-NOT gate,\be
\proj{0}\otimes I + \proj{1}\otimes \sigma_x,
\ee
permutes the $\ket{0}$ and $\ket{1}$ states of the target (second qubit) only if the control (first qubit) is in state $\ket{1}$, it is easy to check that the two-qubit state is transformed according to
\be
\ket{d}\otimes\ket{\alpha} \stackrel{\mbox{C-NOT}}{\longrightarrow} \frac{1}{\sqrt{2}}(U_{\alpha}\ket{d}\otimes\ket{0} + U_{\alpha}^{\dagger}\ket{d}\otimes \ket{1}).
\ee
 Therefore, a projective measurement in the $\{\ket{0}, \ket{1}\}$ basis of the program register will make the data qubit collapse either into the desired state $U_{\alpha}\ket{d}$ or into the wrong state $U_{\alpha}^{\dagger}\ket{d}$, with each outcome having prior probability $1/2$. That is, we have already constructed a probabilistic programmable quantum gate with error rate $\epsilon = 1/2$ (see figure (\ref{fig1})). Notice that a single qubit has been sufficient for Alice to store an arbitrary unitary $U_{\alpha}$, i.e., one from an infinite set, although its recovery only succeeds with probability $1/2$. If Bob obtains $U_{\alpha}^{\dagger}\ket{d}$ instead of $U_{\alpha}\ket{d}$, then not only he fails at performing the wished operation, but in addition he does no longer have the initial data state $\ket{d}$. 

\subsection{Multi-qubit programs.}

\begin{figure}
 \epsfysize=2.2cm
\begin{center}
 \epsffile{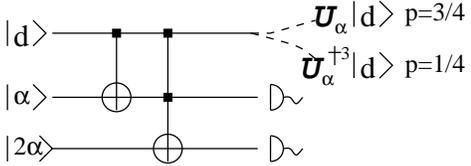}
\end{center}
 \caption{ The gate of figure (\ref{fig1}) can be improved by making a conditional correction of the output after its C-NOT gate. This is achieved by means of a Toffoli gate, which acts as a C-NOT between the first and third line of the circuit only when the second line carries a $\ket{1}$, which corresponds to a failure in the circuit of figure (\ref{fig1}). Bob can measure the second and third lines of the circuit. Only if he obtains $1$ for both outcomes (which happens only one forth of the times, $\epsilon=1/4$) is the transformation unsuccessful. \label{fig2}}
\end{figure}

How can we construct a more efficient programmable gate? Notice that in case of failure, a second go of the previous gate can correct $U_{\alpha}^{\dagger}\ket{d}$ into $U_{\alpha}\ket{d}$. Indeed, Bob needs only apply the gate of fig. (\ref{fig1}) to $U_{\alpha}^{\dagger}\ket{d}$, inserting a new program state, namely $\ket{2\alpha}$, which Alice can prepare by performing twice the operation $U_{\alpha}$ on $(\ket{0}+\ket{1})/\sqrt{2}$. Therefore, if Alice supplies the state $\ket{\alpha}\otimes\ket{2\alpha}$ to Bob, he can perform the operation $U_{\alpha}$ with probability $3/4$. Figure (\ref{fig2}) displays a more compact version of this second probabilistic programmable gate, which requires a two-qubit program register and has a probability of failure $\epsilon=1/4$. 

\begin{figure}
 \epsfysize=3cm
\begin{center}
 \epsffile{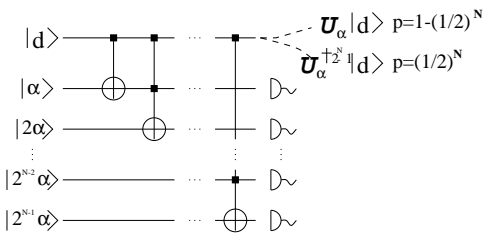}
\end{center}
 \caption{This probabilistic programmable quantum gate uses a $N$-qubit program register and succeeds with probability $p = 1-(1/2)^{N}$, i.e. $\epsilon = (1/2)^{N}$. If no final measurement on the program register is made, or its result is ignored, then this circuit can be regarded as an approximate programmable quantum gate with performance fidelity $F\geq 1-(1/2)^{N}$. 
\label{fig3}}
\end{figure}

 In case of a new failure, the state of the system becomes $U_{\alpha}^{\dagger 3}\ket{d}$. Bob can insert again this state, together with state $\ket{4\alpha}$, into the elementary gate. If Bob has no luck and keeps on obtaining failures, he can try to correct the state as many times as he wishes, provided that the state $\ket{2^{l}\alpha}$ is available for the $l$th attempt. Therefore, for any $N$, the $N$-qubit state $\otimes_{l=1}^{N}\ket{2^l\alpha}$ can be used to implement the transformation $U_{\alpha}$ with probability $1- (1/2)^N$.\footnote{
Note that our several-step correcting scheme for implementing $U_{\alpha}$ resembles that used in \cite{Cirac00} to implement a non-local unitary operation. In the present context all intermediate measurements and conditional actions can be substituted by a single unitary operation, as described in Figures 2 and 3. In this section we have first presented the several-measurement version for pedagogical reasons.
}
 The corresponding probabilistic programmable gate (see figure (\ref{fig3})), consists of the unitary transformation of $\ket{d}\otimes(\otimes_{l=1}^{N}\ket{2^l\alpha})$ into 
\be
\frac{1}{2^{N/2}}(\sqrt{2^N\!-\!1}~U_{\alpha}\ket{d}\otimes\ket{\mbox{r}} + U_{\alpha}^{(2^N\!-\!1)\dagger}\ket{d}\otimes \ket{\mbox{w}})
\label{Nqubits}
\ee
and of a posterior measurement of the program register (either in state $\ket{\mbox{r}}$ or $\ket{\mbox{w}}$, $\braket{\mbox{r}}{\mbox{w}}=0$). Its failure probability, $\epsilon = (1/2)^{N}$, decreases exponentially with the size $N$ of the program register. 

It is interesting to look at how long the program needs to be, on average, until Bob succeeds to perform $U_{\alpha}$ with certainty. With probability $p_1=1/2$ he succeeds after using a single-qubit program; with probability $p_2=1/4$ a two-qubit program is sufficient; etc. The average length $\bar{N}$ of the required program is thus
\be
\bar{N}= \sum_{N=1}^{\infty} p_N N = \sum_{N=1}^{\infty} \frac{N}{2^N} = 2.
\label{average}
\ee
That is, a two qubit register is sufficient, on average, to store an arbitrary $U_{\alpha}$ so that it can be performed with certainty.

\subsection{Probabilistic versus approximate programmable gates.}

A probabilistic programmable gate may either succeed or fail, depending on the result of the final measurement on the program register. An approximate gate, instead, performs a transformation only similar to the desired one, but it is always successful. Suppose we want to apply the unitary transformation $U$ on $\ket{\psi}$ but instead another (general) transformation $\E$ is actually performed. A possible way of quantifying how similar these two operations are is by applying both operations to the same state $\ket{\psi}$, and then computing the fidelity between the two transformed states, $\E(\proj{\psi})$ and $U\ket{\psi}$. When averaged over all possible $\ket{\psi}$ this reads
\bea
F(\E,U) \equiv \int d\psi \bra{\psi}U^{\dagger}\E(\proj{\psi})U\ket{\psi}. 
\label{fid}
\eea
Suppose now that after the transformation (\ref{Nqubits}) of the previous probabilistic gate we decide to ignore the state of the $N$-qubit program register. Then the programmable gate works approximately, implementing an operation $\E_U(\rho)\equiv(1-\epsilon)U_{\alpha}\rho U_{\alpha}^{\dagger}+\epsilon \tilde{U}\rho\tilde{U}^{\dagger}$, where $\tilde{U}\equiv  U_{(1\!-\!2^N)\alpha}$. The average fidelity of performance (\ref{fid}) satisfies $F(\E_U,U) \geq 1-\epsilon = 1-(1/2)^N$.

\section{Programable gates and secret computation on a public computer}

So far we have explicitly constructed programmable quantum gates that perform, either probabilistically or approximately, some class of one-qubit unitary operations $U_{\alpha}$. But the previous protocols also allow Alice to codify with finite resources any unitary operation $V^{(l)}$ acting on an arbitrary number $l$ of qubits. Indeed, as already mentioned, Alice can codify an arbitrary one-qubit unitary operation using only $3$ $U_{\alpha}$'s, and then also combine several of those with C-NOT gates to obtain $V^{(l)}$. 

\subsection{Secret computation on a public computer?}
In view of these results, one may wonder whether quasi-perfect programmable gates can be applied, in the context of quantum cryptography and computation, to secretly compute some unitary operation $V^{(l)}$, for instance a precious algorithm, on some initial $l$-qubit state $\ket{d^{(l)}}$. The idea is that Alice gives a program state $\ket{\P_{V^{(l)}}}$ and the data state $\ket{d^{(l)}}$ to Bob, who operates a programmable quantum gate array but ignores $V^{(l)}$. Bob is required to compute $V^{(l)}\ket{d^{(l)}}$, but Alice does not want Bob to know what program he is running on his quantum computer. If the gate is perfect as in (\ref{perfect}), then Bob can in principle distinguish $\ket{\P_{V^{(l)}}}$ from any other program state, since they are orthogonal. Therefore he can, imperceptibly to Alice, make an illegal copy of the program, perform the required transformation using the original program state, and give the computed state to Alice. 

However, when the gate is slightly imperfect, different programs need no longer be orthogonal. Now Bob can not determine perfectly well which program he is to run in his computer. If he tries to estimate $\ket{\P_{V^{(l)}}}$, then in addition he will necessarily modify the program state, which will result in an improper performance of the gate and then Alice---who may have simply been testing Bob's integrity---can, in principle, detect it. That is, it is not possible for Bob to copy, even in an approximate form, the program state and at the same time perform the operation Alice has commended him with, without this being detectable. 

We next derive lower bounds on the size $N$ of the program register of any quasi-perfect (i.e. with $\epsilon \rightarrow 0$) programmable gate, and on the degree of orthogonality between its program states $\ket{\P_U}$ corresponding to similar operations, in terms of its failure parameter $\epsilon$. These bounds represent a severe limitation on the degree of reliability that a security scheme based on the above ideas can offer. They indicate that the program vectors $\ket{\P_U}$ and $\ket{\P_V}$ are significantly non-orthogonal (that is, non-distinguishable) only when the imperfection parameter $\epsilon$ makes them effectively equivalent.

\subsection{Upper bound to the indistinguishability of states for different programs.}

Let us consider a generic imperfect programmable gate acting on a $\C^n$ system, so that it can be programmed to perform some or all $U\in SU(n)$. It can be described by a unitary operator $G_{\epsilon}$ according to
\be
G_{\epsilon}[\ket{d}\otimes\ket{\P_U}] = \sqrt{1\!-\!\epsilon}(U\ket{d})\otimes\ket{\R^d_U} + \sqrt{\epsilon} \ket{\w^{d}_U},
\label{rotllo}
\ee
where the wrong state $\ket{\w^{d}_U}$ is not required to fulfill any requirement for an {\em approximate} gate, whereas it must satisfy $\braket{\R^{d_1}_{V}}{\w^{d_2}_U}=0$ (here $0$ is the null vector of the data register) for any two inequivalent operations $V$ and $U$ and any two data states $\ket{d_1}$ and $\ket{d_2}$ for a {\em probabilistic} gate. This last condition is necessary for Bob, who ignores both $U$ and the data, to be able to know whether the transformation $U$ has been successfully performed by measuring the program register. 

We first notice that the state $\ket{\R^d_U}$ only depends on $d$ through a contribution of order $\epsilon^{\tau}$, where from now on $\tau=1/2$ for approximate gates and $\tau=1$ for probabilistic ones. Indeed, for any  program state $\ket{\P_U}$, the scalar product of (\ref{rotllo}) corresponding to any two data states $\ket{d_1}$ and $\ket{d_2}$ reads
\be
\braket{d_1}{d_2} = \braket{d_1}{d_2}\braket{\R^{d_1}_U}{\R^{d_2}_U} + O(\epsilon^p),
\ee
[$O(\epsilon^{\tau})$ is a term linear in $\epsilon^{\tau}$] from which, by fixing $\ket{d_1}$ and considering any $\ket{d_2}$, we find that $\braket{\R^{d_1}_U}{\R^{d_2}_U} = 1 + O(\epsilon^{\tau})$. That is 
\be
\ket{\R^d_U} = \ket{\R_U} + O(\epsilon^{\tau}).
\ee
Keeping this in mind, we now consider, for any given $\ket{d}$, the scalar product of (\ref{rotllo}) corresponding to two unitary operations $U$ and $V$, which turns out to read
\be
\braket{\P_U}{\P_V} = \bra{d}U^{\dagger}V\ket{d}\braket{R_U}{R_V} + O(\epsilon^{\tau}).
\label{productUV}
\ee
The scalar product $\braket{\P_U}{\P_V}$ does not depend on $\ket{d}$. Therefore the dependence of $\bra{d}U^{\dagger}V\ket{d}\braket{R_U}{R_V}$ on $\ket{d}$ has to be of order $\epsilon^{\tau}$, at most. Suppose $U$ and $V$ are very close. That is,
\be
U^{\dagger} V = e^{iL} = I + iL -\frac{1}{2} L^2 + O(L^3),
\ee
where all the eigenvalues $\alpha_1 \geq \alpha_2 \geq ...\geq \alpha_n$ of the traceless ($\sum_i \alpha_i=0$) hermitian operator $L=$ $\sum_i^{n} \alpha_i\proj{i}$, are very small. The largest variation of $\bra{d}U^{\dagger}V\ket{d}$ in (\ref{productUV}) for two different vectors $\ket{d}$ is $\bra{1}U^{\dagger}V\ket{1}-\bra{n}U^{\dagger}V\ket{n}=\alpha_1+|\alpha_n|$. We introduce a distance on the set of operators on $C^n$,
\be
D_{U,V} \equiv (\mbox{Tr} [(U-V)^{\dagger}(U-V)])^{\frac{1}{2}}.
\ee
Then $\alpha_1 +|\alpha_n| \geq (\sum_i \alpha_i^2)^{1/2}/n= D_{U, V}/\sqrt{2}n$. Subtracting (\ref{productUV}) for $\ket{d}=\ket{n}$ from itself for $\ket{d}=\ket{1}$ we conclude that $|\braket{\R_U}{\R_V}| \leq O(\epsilon^{\tau})n/D_{U,V}$, which finally implies 
\be
|\braket{\P_U}{\P_V}| \leq \frac{O(\epsilon^{p})n}{D_{U,V}}.
\label{bound1}
\ee
This bound says that in a programmable quantum gate with a small error rate $\epsilon \ll 1$, two transformations $U$ and $V \in SU(n)$ will have program states with significant overlap $\braket{\P_U}{\P_V}$ (states $\ket{\P_U}$ and $\ket{\P_V}$ are indistinguishable) only if $U$ and $V$ are also very close to each other, $D_{U, V} \ll 1$. That is, only if $U$ and $V$ process the data very similarly, then a dishonest Bob is unable to distinguish between the corresponding programs. 

\subsection{Lower bound to the dimension of the program register.}

 The previous result can also be used to derive a lower bound on the dimension of the program register of an imperfect programmable gate with error $\epsilon$. For simplicity we will assume that the gate can only be programmed to perform the one-qubit transformations $U_{\alpha}$ from eq. (\ref{unialpha}). Consider a discrete subset of such transformations, namely those with $\alpha_s \equiv \pi s/M$, $s=0 ,..., M-1$, and apply the previous bound to $U_{\alpha_s}$ and $U_{\alpha_{s+1}}$. We obtain 
\be
|\braket{\P_{\alpha_s}}{\P_{\alpha_{s+t}}}| \leq K\epsilon^{p}M,~~~~ \forall t\neq 0,
\label{bo}
\ee
where $K$ is some unimportant constant. We need the following lemma.

\vspace{.2cm}

{\bf Lemma:} Let $\{\ket{\psi_i}\}_{i=1}^q$ be a set of $q$ (normalized) vectors such that their scalar products $\nu_{ij}\equiv \braket{\psi_i}{\psi_j}$ satisfy $|\nu_{ij}| < q^{-1}$ for $i\neq j$. Then the $q$ vectors are linearly independent.
 
\vspace{.2cm}

\noindent{\em Proof:} The rank of the set $\{\ket{\psi_i}\}_{i=1}^q$ is equal to the rank of the matrix $N$, $N_{ij}\equiv \braket{\psi_i}{\psi_j}$, which has ones in all diagonal entries. The modulus of any entry of the matrix $N\!-\!I$ is smaller than $q^{-1}$. Let $\ket{\varphi}$ be a normalized eigenvector of $N\!-\!I$, with eigenvalue $\lambda$. Then $\lambda[\ket{\varphi}]_i=[(N\!-\!I)\ket{\varphi}]_i = \sum_{j=1}^q \nu_{ij} [\ket{\varphi}]_j$, where $[~]_j$ denotes the $j$th vector component. Let $i$ be such that $|[\ket{\varphi}]_i| \geq |[\ket{\varphi}]_j|$ $\forall j$. Then $|\sum_{j=1}^q \nu_{ij} [\ket{\varphi}]_j| \leq \sum_{j=1}^q |\nu_{ij}||[\ket{\varphi}]_i|$ $<$ $|[\ket{\varphi}]_i|$, that is, $|\lambda| < 1$, and since this holds for all the eigenvalues of $N-I$, $N$ has $q$ positive eigenvalues or, equivalently, rank $q$.

\vspace{.2cm}

{\bf Remark:} For $q$ sufficiently large, if all $\nu_{ij}$ are of order $\nu$ and the components of the eigenvector $\ket{\varphi}$ are relatively equally weighted, then it is plausible that the eigenvalue $\lambda = \sum_{j=1}^q \nu_{ij} [\ket{\varphi}]_j/[\ket{\varphi}]_i$ is of the order $\nu q^{1/2}$ (random walk). This suggests that in order for the set $\{\ket{\psi_i}\}_{i=1}^q$ to have rank close to $q$, it is sufficient that $|\nu_{ij}| < q^{-1/2}$, instead of $|\nu_{ij}| < q^{-1}$ as required in the lemma.

\vspace{.2cm}

Let us set $M\equiv (\epsilon^{\tau}K)^{-1/2}$. Then (\ref{bo}) becomes $|\braket{\P_{\alpha_s}}{\P_{\alpha_{s+t}}}|\leq 1/M$, and this means, because of the lemma, that at least an $M$-dimensional Hilbert is required to contain $\{\ket{\P_{\alpha_s}}\}_{s=0}^{M-1}$. That is, the program register must consists of at least $(\tau/2)\log(1/\epsilon)$ qubits. Notice that the previous remark suggests that this bound may be reduced to $\tau\log(1/\epsilon)$ qubits, in which case the probabilistic programmable gate of figure (\ref{fig3}) would require, asymptotically, the smallest possible program register. For a general programmable gate implementing some or all transformations $U\in SU(n)$ it is straightforward to obtain a similar lower bound on the dimensions of the program register, which also says that its number $N$ of qubits grows proportionally to the logarithm of the inverse of the rate error, $N=k_n\log (1/\epsilon)$, for some positive constant $k_n$. 

\section{Applications: Manipulation of unknown quantum dynamics}

 We have shown how to store an arbitrary unitary transformation in the pure state of a finite quantum register, in such a way that it can be performed quasi-perfectly at a later time. Once the unknown operation has been encoded in a quantum state, it can of course be processed using {\em any} known state manipulation technique.

\subsection{Unidirectional quantum remote control.}

 An interesting application of our results is in the context of quantum remote control. As introduced by Huelga {\em et al.} in \cite{Hue00}, let us suppose Bob wishes to manipulate some data state according to an unknown operation Alice, a distant party, can implement by using some device. If the state of Alice's device cannot be teleported, then the optimal protocol \cite{Hue00} is to use standard teleportation \cite{tele} to send the data from Bob to Alice, who will use the device to process it and will teleport it back to Bob.

 But we now know how to efficiently store operations in quantum states, which can then be teleported. This leads to a new scheme for quantum remote control: Alice stores the operation in a quantum state and applies standard teleportation to send it to Bob. 

 Remarkably enough, in this protocol only one-way communication is required ---in addition to entanglement ---in order for Alice to remotely manipulate Bob's data, as opposed to the two-way classical communication of the scheme presented in \cite{Hue00}. This implies that the operation can be teleported independently of whether Bob's data state is already available. More specifically, we find that a $N$-qubit program $\ket{\alpha}\otimes ... \otimes \ket{2^N \alpha}$ can be teleported from Alice to Bob by using up $N$ ebits of entanglement and by sending $N$ classical bits from Alice to Bob (recall that the classical communication cost of quantum teleportation of equatorial states, as $\ket{\alpha}$, require only one bit per state \cite{lo}). Eq. (\ref{average}) implies that, on average, $2$ ebits of entanglement between Alice and Bob, and $2$ classical bits from Alice to Bob are sufficient for Alice to teleport an arbitrary $U_{\alpha}$ to Bob, so that he, ignoring $U_{\alpha}$, can perform it with certainty.

\subsection{Estimation of quantum dynamics and storage of non-local transformations.}

 The storage of quantum transformations turns out to be useful in several other contexts. If state estimation techniques are applied to the system that stores an unknown operation, then we obtain the scheme for estimation of quantum dynamics recently exploited by Ac\'{\i}n {\em et al} \cite{acin}.

 Cirac {\em et al} \cite{Cirac00} have recently explored the possibilities of encoding operations in quantum states in the context of non-local transformations of a composite system. In particular, they have shown how to implement non-local unitary transformations using less than one ebit of entanglement.  In an extension of their work, D\"ur {\em et al} \cite{Wolf} have considered alternative schemes for storing and manipulating quantum transformations.

\section{Conclusions}

We have presented a scheme for storing unitary operations in the quantum state of a finite dimensional program register. The operations can be implemented at a later time with some associated error $\epsilon$, which decreases exponentially with the number of qubits of the program register. We have presented both probabilistic and approximate programmable quantum gates, and have discussed the possibility of using them to make a secrete computation on a public quantum computer. Finally, a unidirectional scheme for remote manipulation of quantum states has also been put forward. 

\section*{acknowledgments}

 We thank W. D\"ur for useful comments. G.V. acknowledges a Marie Curie Fellowship (HPMF-CT-1999-00200, European Community). This work was also supported by the SFB project 11 on ``control and measurement of coherent quantum systems'' (Austrian Science Foundation), the Institute for Quantum Information GmbH and the project EQUIP (contract IST-1999-11053, European Community).

\begin{references}
\bibitem{explanation} Indeed, an arbitrary unitary operation $V^{(l)}$ on $l$ qubits can decomposed as a fixed ---that is, $V^{(l)}$-independent--- sequence of ($V^{(l)}$-dependent) one-qubit unitary operations and a specific kind of two-qubit unitary operations, for instance C-NOT gates ---since the sequence is fixed, $V^{(l)}$ is specified by the one-qubit transformations only; an arbitrary von Neumann measurement, given by orthogonal projectors $\{P_k\}$, can be decomposed into a general unitary operation $V$, with $VP_kV^{\dagger}\equiv Q_k$, a von Neumann measurement on a fixed set of projectors $\{Q_k\}$ and a second unitary operation $V^{-1}$. Finally, a general quantum operation on an arbitrary system can be achieved by appending an ancilla in some blank state to the system, performing a unitary transformation $U$ followed by a von Neumann measurement, $\{P_k\}$, and disposing of part of the system and/or ancilla. Therefore any possible transformation of a quantum system can be essentially specified by means of one-qubit unitary operations.
\end{references}
\end{document}